\documentclass[aps,pre,twocolumn,superscriptaddress]{revtex4-2}
\usepackage{mathtools,amssymb}
\usepackage{graphicx}
\usepackage{epstopdf}
\begin{document}

\title{Fluctuation theorems for a non-Gaussian system}
\author{A. Saravanan}
\affiliation{Department of Physics, St Joseph's University,\\ Bengaluru-560027, Karnataka, India.}
\author{I. Iyyappan}
\email[email: ]{iyyap.si@gmail.com}
\affiliation{Division of Sciences, Krea University,\\ Sri City-517646, Andhra Pradesh, India.}
\begin{abstract}
In this work, we numerically investigate the Jarzynski equality and the Crooks fluctuation theorem for a Brownian particle diffusing in a heterogeneous thermal bath, which gives rise to a non-Gaussian position distribution. The thermal bath heterogeneity is modeled using the diffusing-diffusivity concept, where the mobility is considered as a fluctuating quantity. We confine the Brownian particle by a time-dependent harmonic potential, and by changing the stiffness coefficient, an isothermal process can be performed. Using the framework of stochastic thermodynamics, the stochastic work was calculated. Our results indicate that the Jarzynski equality and the Crooks fluctuation theorem remain valid for a non-Gaussian system. Furthermore, we also find that the work distribution is non-Gaussian for the diffusing-diffusivity system even for long process times.

\end{abstract}

\maketitle
\section{Introduction} 
Classical thermodynamics is one of the most elegant and universal theories of physics which can be applied to a large class of systems from Brownian particles to black holes. The initial motivation of thermodynamics studies was focused on improving the steam engine efficiency, which converts the heat into mechanical work. The second law of thermodynamics imposes an upper bound on the extractable work (W) from the thermal bath, $W\geq \Delta F$ \cite{clausius1879mechanical}. Here, $\Delta F$ is the Helmholtz free-energy difference. Equality is attained only for the reversible process. For a macroscopic system having $10^{23}$ degrees of freedom, fluctuations in thermodynamic quantities such as heat, work, internal energy, etc., are negligible \cite{jarzynski2012equalities}. In contrast, if the system contains only a few degrees of freedom, then the fluctuations become significant \cite{bustamante2005nonequilibrium}. In this context, Sekimoto showed that the thermodynamic quantities such as heat and work can be computed for a Brownian particle at single trajectory level \cite{sekimoto1997kinetic}. Nevertheless, physical quantities can be well described universally either by probability distributions or ensemble averages \cite{jarzynski2012equalities}. 

Over twenty-five years ago, Jarzynski identified the remarkable relation between the equilibrium (Helmholtz) free-energy difference and the exponential average of work performed during the non-equilibrium process \cite{jarzynski1997nonequilibrium}
\begin{equation}\label{a}
\langle e^{-\beta w}\rangle=e^{-\beta\Delta F},
\end{equation}
where $\beta=(\kappa_B T)^{-1}$. $\kappa_B$ is the Boltzmann constant, and $T$ is the bath temperature. $w$ is the stochastic work performed under a non-equilibrium process. The free-energy difference, $\Delta F\equiv F_f-F_i$. Here, $F_{i(f)}$ is the initial (final) Helmholtz free-energy. $F_j=-\kappa_B T\ln Z_j$. The partition function, $Z_j=\int_{-\infty}^{\infty}dx \exp[-\beta V_j(x)]$. Subsequently, Crooks derived the relation between the probability distribution of work in the forward process and the negative of work distribution in the time-reversal process \cite{crooks1998nonequilibrium,crooks1999entropy}
\begin{equation}\label{b}
\frac{P_F(w)}{P_R(-w)}=e^{\beta (w-\Delta F)}.
\end{equation}
Here, $P_F(w)$ is the probability distribution of work during the forward driving protocol $\lambda_F(t)$ for the time $\tau$.  $P_R(-w)$ is the negative of work distribution obtained from the time-reversal protocol $\lambda_R(t)=\lambda_F(\tau-t)$, and hence the forward and reverse processes have the same driving speed. Jarzynski equality and Crooks fluctuation theorem have been thoroughly verified in experiments \cite{liphardt2002equilibrium,carberry2004fluctuations,collin2005verification,douarche2005experimental,douarche2006work,harris2007experimental,saira2012test,martinez2013effective} as well as numerous theoretical studies \cite{jarzynski1997equilibrium,searles1999fluctuation,darve2001calculating,gore2003bias,pohorille2010good,speck2011work,lacoste2015stochastic,marathe2019convergence,dabelow2019irreversibility}. Further, a pedagogical experimental demonstration of Jarzynski equality and Crooks fluctuation for a Brownian particle confined by an optical tweezer was presented by Martins \textit{et. al.}   \cite{martins2025brief}.

There has been ongoing debate in the literature regarding the validity of the Jarzysnki equality, and Crooks fluctuation theorem for non-Gaussian systems \cite{cohen2004note,jarzynski2004nonequilibrium}. Blickle et. al. showed that the fluctuation theorems are valid for a non-Gaussian system in which the non-Gaussian distribution position originates from the nonharmonic potential \cite{blickle2006thermodynamics}. In contrast, in the presemt work, we consider the Brownian particle confined by a harmonic potential and the non-Gaussian distribution of the position arises from heterogeneity in the thermal bath \cite{wang2009anomalous,toyota2011non,wang2012brownian,luo2018non,chakraborty2020disorder,pastore2021rapid}. Such non-Gaussian distributions of position are well described by the diffusing-diffusivity model  \cite{chubynsky2014diffusing,chechkin2017brownian}. Despite the non-Gaussian position distribution, these systems show normal diffusion. Therefore, it is generally called a Brownian (Fickian) yet non-Gaussian diffusion. 

Recent studies have shown that non-Gaussianity can reduce the stochastic heat engine performance \cite{iyyappan2023brownian}. On the other hand, non-Gaussianity becomes advantageous when only a small number of searchers are needed to reach the destination \cite{sposini2024being} and becomes a disadvantage when many searchers are needed to reach the destination \cite{sposini2024beingd}. These contrasting roles of non-Gaussianity motivates us to investigate whether the Jarzynski equality and Crooks fluctuation theorem remain valid for such non-Gaussian systems that exhibit normal diffusion.\\ 

\section{Thermal bath with Fluctuating mobility}
We consider a Brownian particle diffusing in a heterogeneous thermal bath \cite{wang2009anomalous,chechkin2017brownian} and  confined by an external potential $V (x,\lambda)$, where $\lambda$ is the time-dependent external control parameter. In the overdamped regime, the one-dimensional Langevin equation is given by 
\begin{equation}\label{d}
\frac{dx}{dt}=-\mu(t)\frac{\partial V(x,\lambda)}{\partial x}+\sqrt{2\mu(t)\kappa_B T}\xi(t),
\end{equation}
\begin{equation}\label{e}
\mu(t)=Y^{2}(t),
\end{equation}
\begin{equation}\label{f}
\frac{dY}{dt}=-\frac{Y}{s}+\sigma\eta(t).
\end{equation}
Here, $x(t)$ denotes the position of the Brownian particle at time $t$, and $\mu(t)$ is the fluctuating mobility. The thermal noise $\xi(t)$ is Gaussian and white, satisfying $\langle\xi(t)\rangle=0$, and $\langle\xi(t_1)\xi(t_2)\rangle=\delta(t_1-t_2)$. The noise $\eta(t)$ accounts for the heterogeneity of the thermal bath which is also Gaussian and white with $\langle\eta(t)\rangle=0$, and $\langle\eta(t_1)\eta(t_2)\rangle=\delta(t_1-t_2)$. The parameter $s$ represents the correlation time of the Orenstein-Uhlembeck process given in Eq. (\ref{f}), and $\sigma$ is the strength of the fluctuation. Eq. (\ref{e}) ensures the positivity of mobility and its average value is given by \cite{iyyappan2023brownian}
\begin{equation}\label{g}
\langle\mu \rangle=\frac{\sigma^{2}s}{2}.
\end{equation}
The non-Gaussianity of the position distribution can be measured by the kurtosis \cite{rahman1964correlations}.
\begin{equation}\label{h}
K_x=\frac{\langle(x-\langle x\rangle)^{4} \rangle}{\sigma_x^{4}}.
\end{equation}
Here, $\sigma_x^{2}=\langle (x-\langle x\rangle)^{2}\rangle$ is the variance of the position. For the Gaussian distribution, $K_x=3$, and if $K_x\neq3$ then the probability distribution deviates from the Gaussian.

The total work performed on the Brownian particle by the external control parameter ($\lambda$) is given by \cite{sekimoto1997kinetic,seifert2012stochastic}
\begin{equation}\label{i}
w=\int_{0}^{\tau}\frac{\partial V(x,\lambda)}{\partial \lambda}\frac{d\lambda} {dt}dt,
\end{equation}
where $\tau$ is the time duration of the process. The work $w$ is a stochastic variable, and when the control parameter $\lambda$ is held constant, then $w$ becomes zero.\\

\section{Breathing parabola}
We consider the harmonic potential with the time-dependent stiffness coefficient 
\begin{equation}\label{j}
V(x,\lambda)=\frac{1}{2}\lambda(t)x^{2}.
\end{equation}
The stiffness coefficient $\lambda(t)$ is varied with time according to a sinusoidal protocol in order to get an (approximately) symmetric work distribution. In particular, we take
\begin{equation}\label{k}
\lambda(t)=A-B\sin\left(\frac{2\pi t}{\tau}\right),
\end{equation}
where $A$, and $B$ are positive constants, and $\tau$ is the period of the driving process. Throughout this work, we set $A=10$, and $B=9$. Since the initial and final values of $\lambda$ are the same, the Helmholtz free-energy difference vanishes, $\Delta F=0$. Therefore, the Jarzyski equality reduces to the simple form
\begin{equation}\label{l}
\langle e^{-\beta w}\rangle=1.
\end{equation}
An analytical verification of the Jarzynski equality and Crooks fluctuation theorem is not feasible for a diffusing-diffusivity model considered in our study. Therefore, we perform the numerical simulation to test the validity of fluctuation theorems for a non-Gaussian system. We use the Euler-Maruyama method to integrate the Langevin equations given in Eqs. (\ref{d})-(\ref{f}). We set the following values: $\kappa_B=1$, $T=1$, and $dt=10^{-5}$ (when $\tau>10$, $dt=10^{-4}$). We consider $10^{6}$ independent trajectories. Since $\lambda(0)=\lambda(\tau)$, the forward and reverse processes share the same initial equilibrium distribution of position. Therefore, for each trajectory, the initial value of position ($x(0)$) is randomly drawn from the same equilibrium distribution. The initial values of $Y(0)$ are chosen from the steady-state solution (see Ref. \cite{chechkin2017brownian,iyyappan2023brownian}). For the diffusing-diffusivity case, we consider two sets of parameters $s$, and $\sigma$ satisfying Eq. (\ref{g}), such average mobility is fixed to $\langle\mu \rangle=1$. For comparison, we also study the constant mobility case described by the following Langevin equation
\begin{equation}\label{}
\frac{dx}{dt}=-\mu\lambda(t)x+\sqrt{2\mu\kappa_B T}\xi(t),
\end{equation}
where the constant mobility is also set to $\mu=1$.

\section{Results and Discussion}
\begin{figure}[hptb]
\centering
\includegraphics[scale=0.3,angle=0]{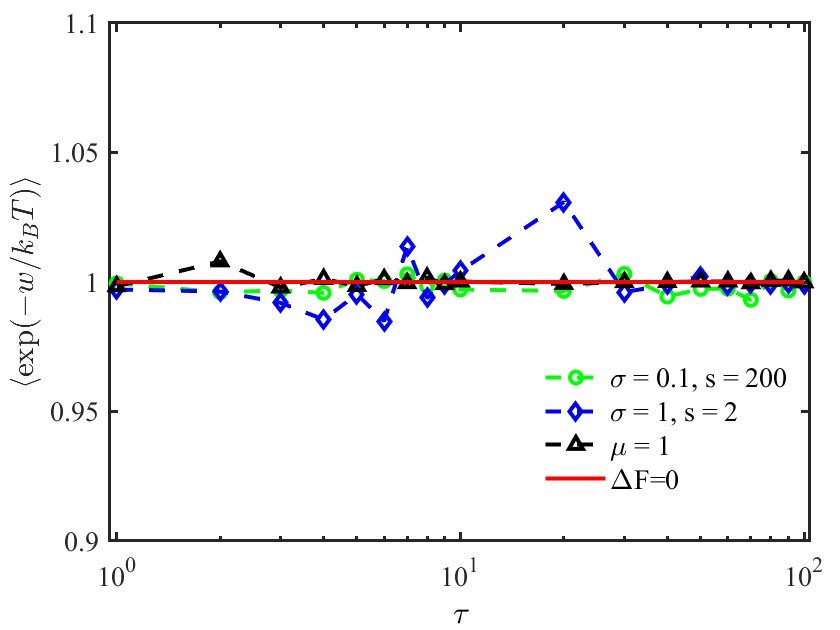}
\vspace{-0.6cm}
\caption{\label{fig:je} The Jarzynski equality in Eq. (\ref{l}) is plotted as a function of process time $\tau$. The black curve corresponds to the Gaussian system with constant mobility $\mu=1$. The blue and green curves represent the non-Gaussian (diffusing-diffusivity) system for the indicated values of $\sigma$, and $s$, chosen such that $\langle\mu\rangle=1$. The red solid line denotes the theoretical prediction, $\langle e^{-\beta w}\rangle=1$.}
\end{figure}

The Jarzynski equality given in Eq. (\ref{l}) is plotted in Fig. (\ref{fig:je}) for different values of process time $\tau$. The results demonstrate that the Jarzynski equality is well satisfied for the systems exhibiting a non-Gaussian distribution of position (blue and green). We only observed the small fluctuation in the $\langle e^{-\beta w}\rangle$ for the diffusing-diffusivity case with $\sigma=1$, and $s=2$, compared with another case (green curve).

\begin{figure}[hptb]
\centering
\includegraphics[scale=0.0748,angle=0]{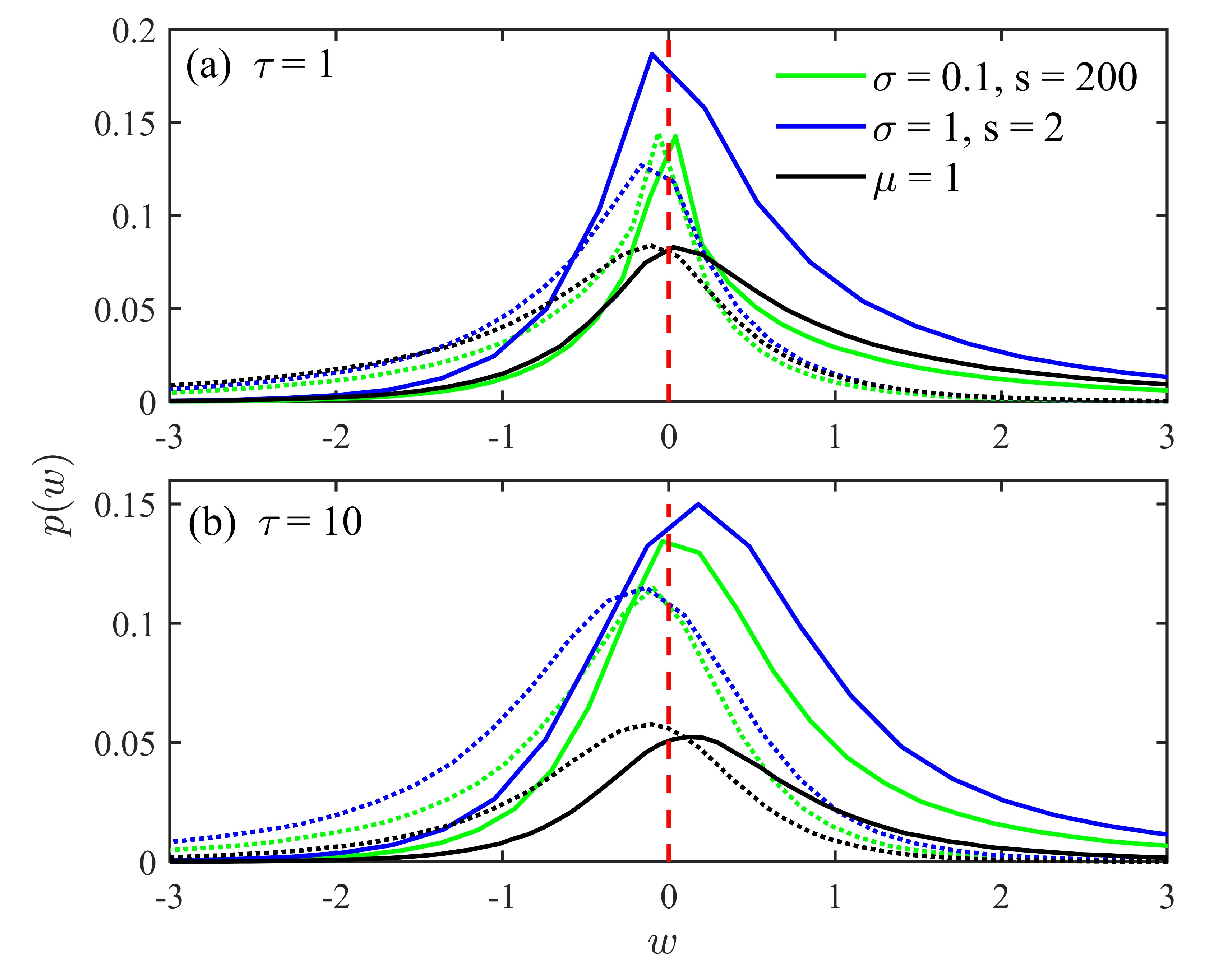}
\vspace{-0.6cm}
\caption{\label{fig:wd} The probability distribution of work is plotted. The solid lines correspond to the forward process while the dotted lines represents the time-reversal process. The black curve denotes the Gaussian system with constant mobility $\mu=1$. The blue and green curves correspond to non-Gaussian (diffusing-diffusivity) systems with indicated values of $\sigma$, and $s$ satisfying $\langle\mu\rangle=1$. The vertical line marks the mean work, $\langle w\rangle=0$.}
\end{figure}

The probability distribution of work for a forward protocol (solid lines) and time-reversal protocol (dotted lines) is shown in Fig. (\ref{fig:wd}). The vertical dashed line denotes the mean work $\langle w\rangle=\Delta F=0$. We found that the probability distribution of forward and reverse work intersects approximately at $\langle w\rangle=0$ for a Gaussian system. However, for the non-Gaussian system, the intersection point is slightly shifted away from $\langle w\rangle=0$, indicating that a larger statistical sampling is required to accurately resolve the crossing for a non-Gaussian system.

\begin{figure}[hptb]
\centering
\includegraphics[scale=0.0748,angle=0]{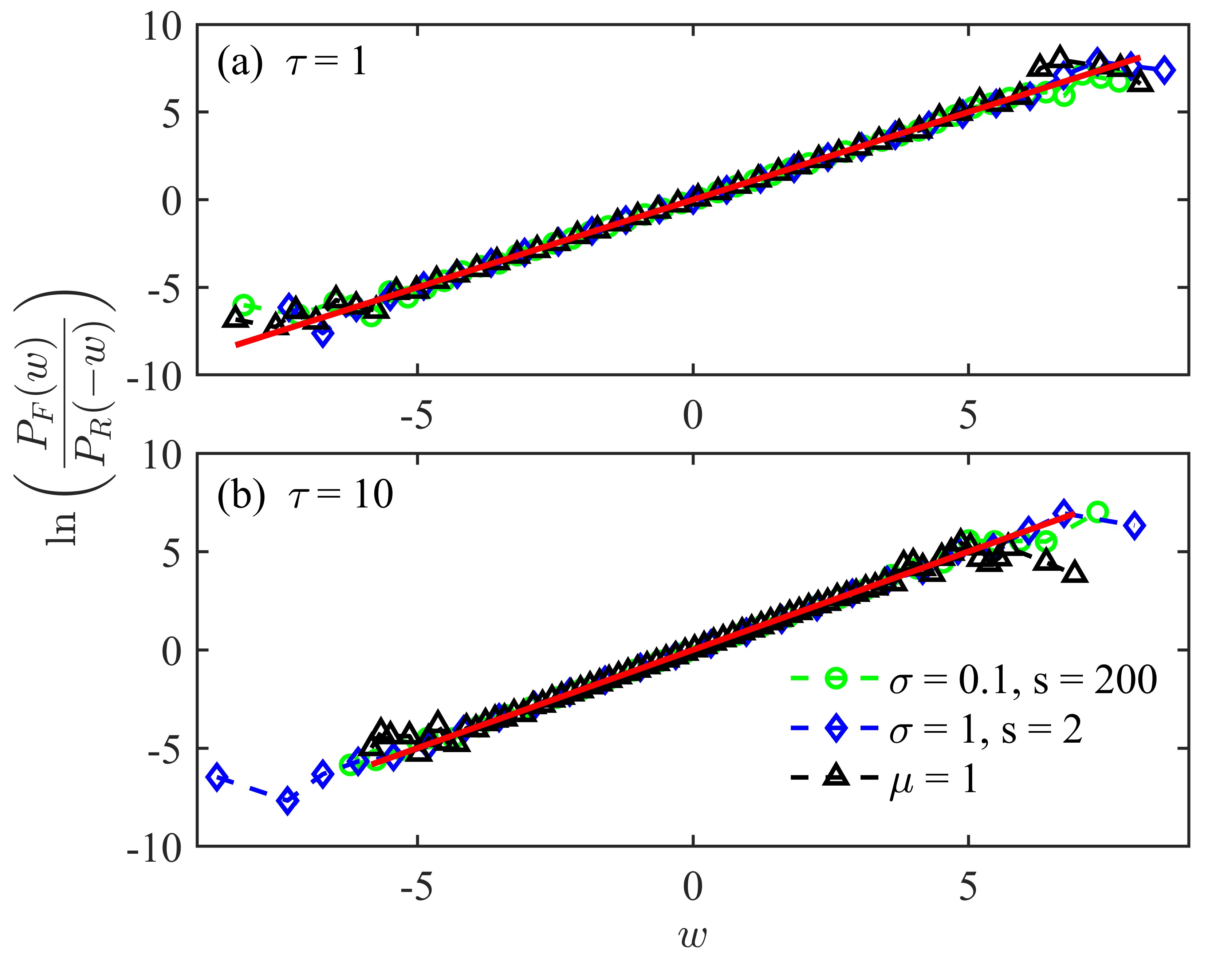}
\vspace{-0.6cm}
\caption{\label{fig:ft} The fluctuation theorem in Eq. (\ref{b}) is plotted for different process times $\tau$. The black curve represents the Gaussian system with constant mobility $\mu=1$. The blue and green curves correspond to a non-Gaussian system with parameters $\sigma$, and $s$ satisfying $\langle\mu\rangle=1$.}
\end{figure}

To verify the Crooks fluctuation theorem given in Eq. (\ref{b}), we plot $\ln\left(P_F(w)/P_R(-w)\right)$ as a function of $w$ in Fig. (\ref{fig:ft}). For both Gaussian and non-Gaussian systems, this quantity exhibits a linear dependence on $w$, and hence the Crook fluctuation theorem is satisfied.

\begin{figure}[hptb]
\centering
\includegraphics[scale=0.075,angle=0]{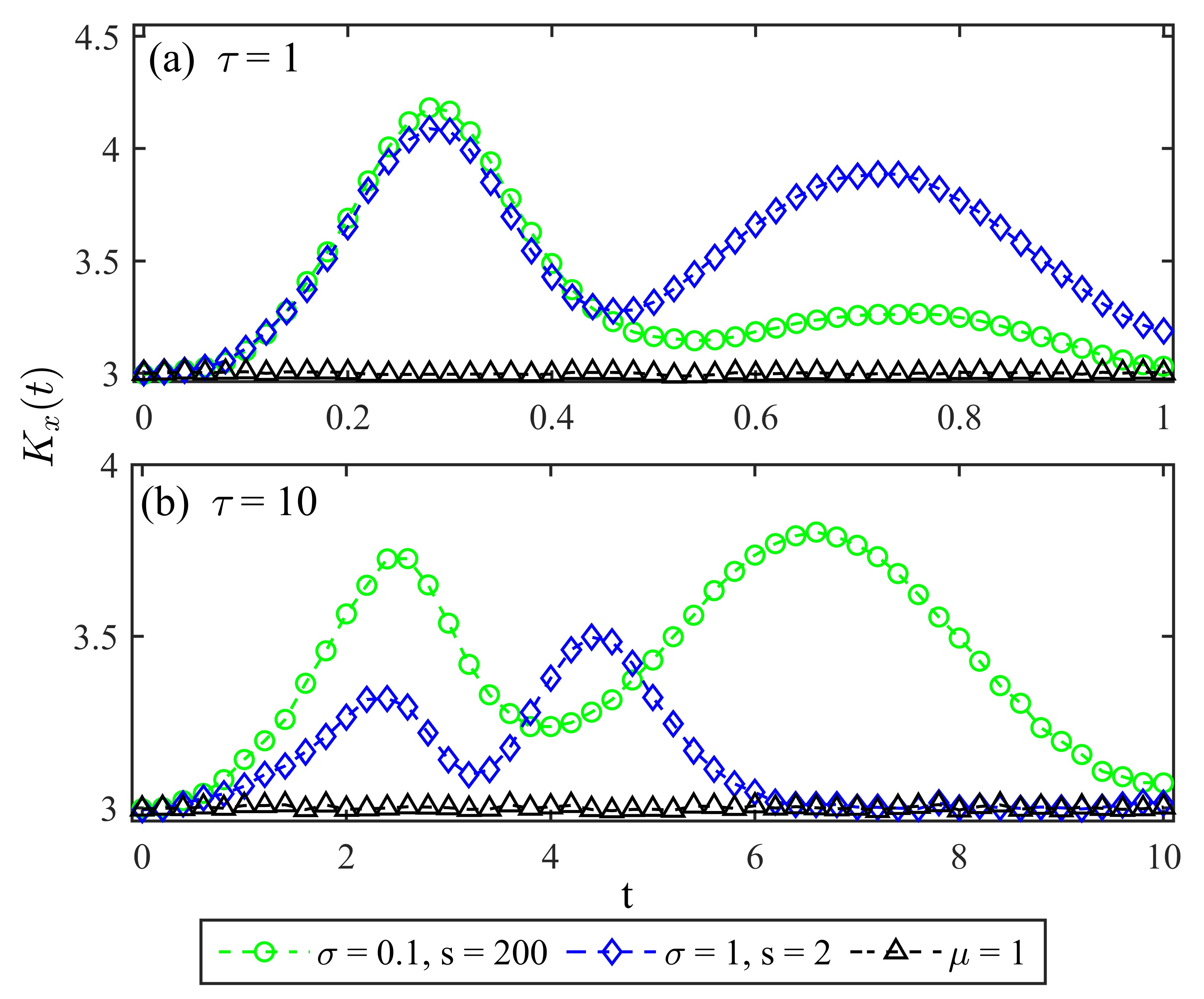}
\vspace{-0.6cm}
\caption{\label{fig:ku} The kurtosis is plotted as a function of process time $\tau$. The blue, and green curves correspond to the non-Gaussian (diffusing-diffusivity) system for the indicated values of $\sigma$, and $s$ satisfying $\langle\mu\rangle=1$. The black curve represents the Gaussian system with $\mu=1$.}
\end{figure}

To demonstrate the presence of non-Gaussianity in the system, we plot the kurtosis of the position as a function of time in Fig. \ref{fig:ku}. Panel (a) corresponds to $\tau=1$, and panel (b) shows results for $\tau=10$. In both cases, the kurtosis accociated with green and blue curves remains higher than $3$, clearly indicating a non-Gaussian position distribution throughout the process.

\begin{figure}[hptb]
	\centering
	\includegraphics[scale=0.0745,angle=0]{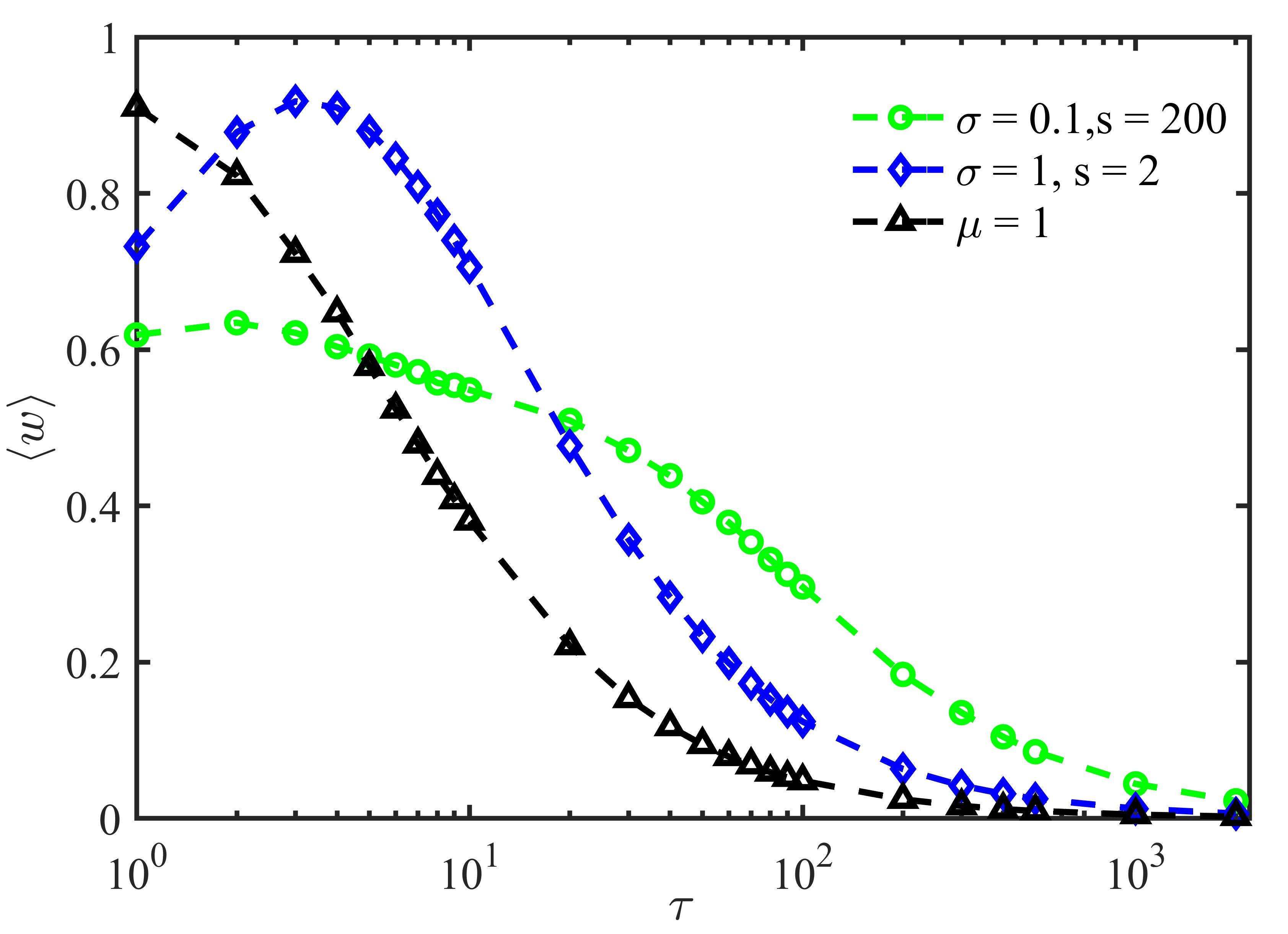}
	\vspace{-0.6cm}
	\caption{\label{fig:Avgw} The average work is plotted as a function of process time $\tau$. The black curve correspond to the Gaussian system with constant mobility $\mu=1$. The blue and green curves represents the non-Gaussian (diffusing-diffusivity) systems for the indicated values of $\sigma$, and $s$ and satisfying $\langle\mu\rangle=1$.}
\end{figure}

In the following analysis, we focus exclusively on the forward isothermal process. Fig. (\ref{fig:Avgw}) shows the average work as a function of the process time $\tau$. For the Gaussian system, the mean work decreases monotonically and approaches zero as $\tau$ increases. However, for the non-Gaussian system, the average work initially increases, attains a maximum at an intermediate process time, and subsequently decreases monotonically.

\begin{figure}[hptb]
	\centering
	\includegraphics[scale=0.0715,angle=0]{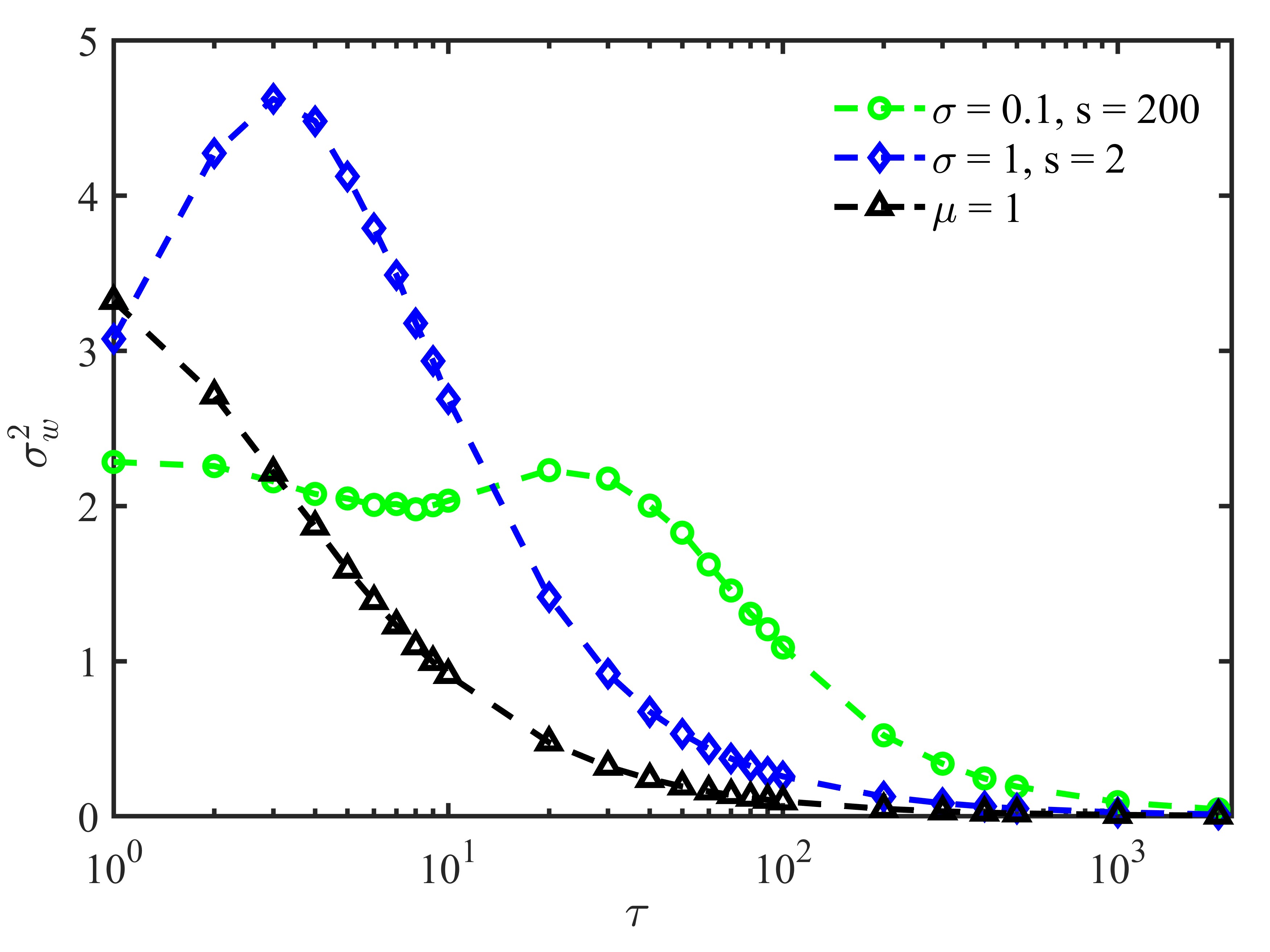}
	\vspace{-0.6cm}
	\caption{\label{fig:Fluw} The variance of work is plotted as a function of the process time $\tau$. The black curve represent the Gaussian system with constant mobility $\mu=1$. The blue and green curves correspond to the non-Gaussian systems with indicated values of $\sigma$, and $s$ and satisfying $\langle\mu\rangle=1$.}
\end{figure}

Next, we analyze the moments of work distribution. Fig. (\ref{fig:Fluw}) shows the variance of work, $\sigma_w^{2}=\langle (w-\langle w\rangle)^{2}\rangle$ as a function of the process time $\tau$. For the Gaussian system, the fluctuation of work decreases monotonically with increasing $\tau$. However, for non-Gaussian systems $\sigma_w^{2}$ exhibits nonmonotonic behaviour with $\tau$ indicating the enhanced fluctuations induced by thermal bath heterogeneity. Overall, non-Gaussianity leads to higher work fluctuations compared to the Gaussian case. In the quasi-static limit ($\tau\rightarrow\infty$), the variance of work vanishes ($\sigma_w^{2}\rightarrow 0$) for both Gaussian and non-Gaussian systems. 
\\

\begin{figure}[hptb]
\centering
\includegraphics[scale=0.0716,angle=0]{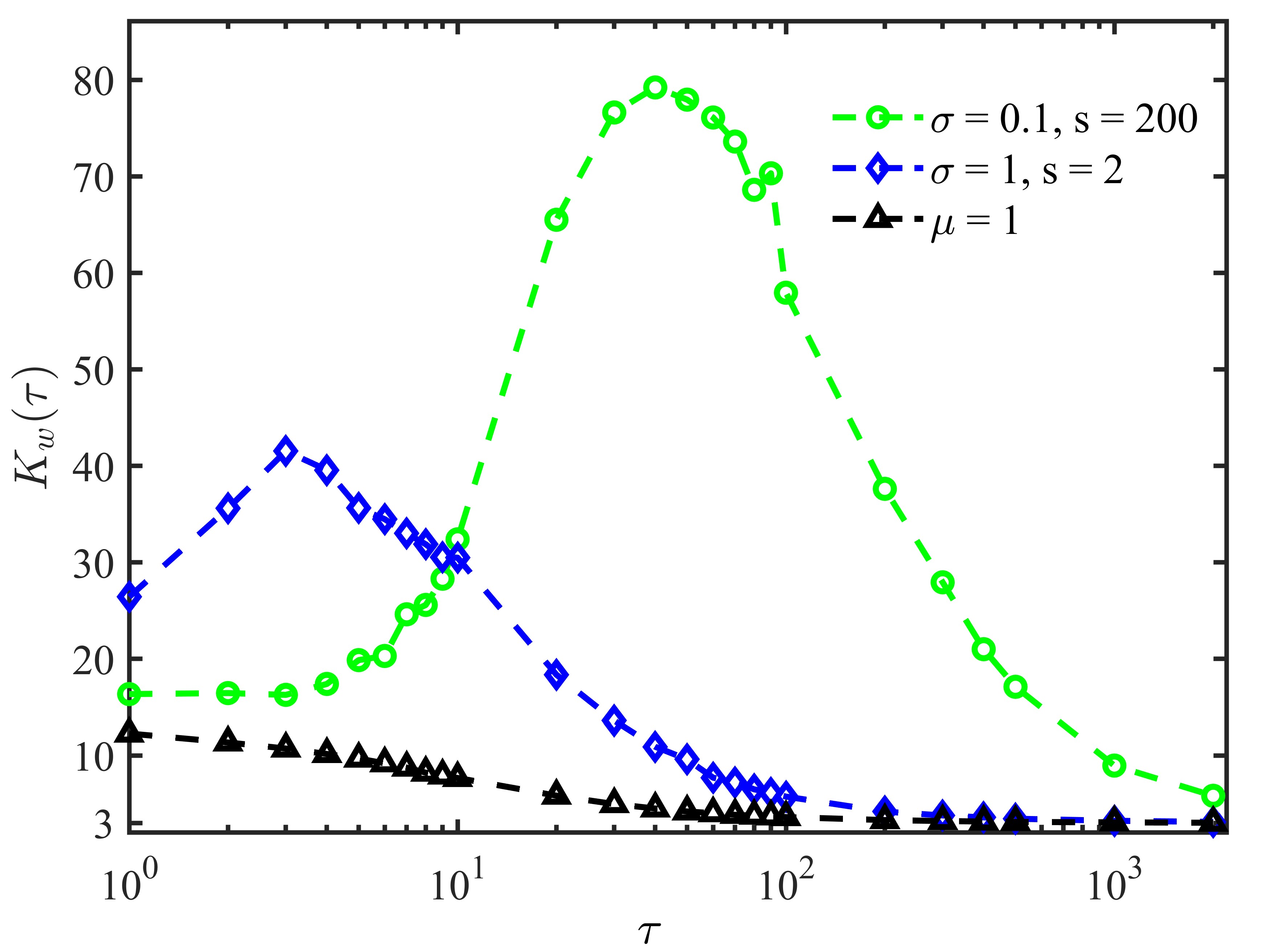}
\vspace{-0.6cm}
\caption{\label{fig:Kuw} The kurtosis of work is plotted as a function of the process time $\tau$. The black curve represent the Gaussian system with constant mobility $\mu=1$. The blue and green curves correspond to the non-Gaussian (diffusing-diffusivity) systems with indicated values of $\sigma$, and $s$, and satisfying $\langle\mu\rangle=1$.}
\end{figure}

Further, we analyze the kurtosis of the work distribution, $K_w=\langle(w-\langle w\rangle)^{4} \rangle/\sigma_w^{4}$, as a function of process time $\tau$, as shown in Fig. (\ref{fig:Kuw}). For a Gaussian system, the work distribution monotonically approaches a Gaussian form around $\tau=100$. In contrast, for a non-Gaussian system, the relation between $K_w$ and the process time $\tau$ is nonmonotonic; $K_w$ initially increases and reaches a maximum than approaches zero at a larger $\tau$. In particularly for $\sigma=0.1$, and $s=200$, the convergence to Gaussian behavior occurs at  much longer times. This delayed relaxation arises from the large correlation time of the fluctuating mobility, which sustains a non-Gaussian position distribution even at long times (see Ref. \cite{iyyappan2023brownian}).  
\\
\vspace{-0.6cm}
\section{Conclusion}
In this article, we studied the Brownian particle diffusing in a heterogeneous thermal bath, which demonstrates the non-Gaussian position distribution at shorter times along with normal diffusive behavior. The dynamics were described using a diffusing-diffusivity model, in which the particle mobility is treated as a fluctuating quantity. Using numerical simulations, we investigated the validity of Jarzynski equality and Crooks work fluctuation theorem for a non-Gaussian system (see Figs. (\ref{fig:je}), (\ref{fig:ft}) and (\ref{fig:ku})). Our resutls shows that both fluctuation theorems remain valid for a non-Gaussian system which shows the robustness of these two results beyond the Gaussian system. Further, analyzing the higher moments of work distribution in the forward process, we show that the work distribution remains non-Gaussian for the diffusing-diffusivity system even for the long process time, highlighting the qualitative difference between the constant diffusivity and diffusing-diffusivity (see Fig. (\ref{fig:Kuw})).\\
\vspace{-0.6cm}
\begin{acknowledgments}
I.I. would like to convey his gratitude to Krea University for financial support through the University's post-doctoral fellowship.
\end{acknowledgments}

\bibliography{ref}
\end{document}